# Cell bystander effect induced by radiofrequency electromagnetic fields and magnetic nanoparticles

G.F. Goya[1,2], L. Asin[1], M. P. Calatayud[1], A. Tres[1,3], M.R. Ibarra[1,2,4*]

[1]Instituto de Nanociencia de Aragón (INA), Universidad de Zaragoza, Mariano Esquillor s/n, 50018- Zaragoza, Spain

[2]Departamento de Física de la Materia Condensada, Facultad de Ciencias, Universidad de Zaragoza, 50009-Zaragoza, Spain

[3]Departamento de Oncología, Hospital Universitario Lozano Blesa, 50009-Zaragoza, Spain

[4]Laboratorio de Microscopías Avanzadas, Universidad de Zaragoza,(LMA), Universidad de Zaragoza, Mariano Esquillor s/n 50018 Zaragoza, Spain,

*Corresponding author: ibarra@unizar.es

**Induced effects by direct exposure to ionizing radiation (IR) are a central issue in many fields like radiation protection, clinic diagnosis and oncological therapies. Direct irradiation at certain doses induce cell death, but similar effects can also occur in cells no directly exposed to IR, a mechanism known as bystander effect. Non-IR (radiofrequency waves) can induce the death of cells loaded with MNPs in a focused oncological therapy known as magnetic hyperthermia. Indirect mechanisms are also able to induce the death of unloaded MNPs cells. Using in vitro cell models, we found that colocalization of the MNPs at the lysosomes and the non-increase of the temperature induces bystander effect under non-IR. Our results provide a landscape in which bystander effects are a more general mechanism, up to now only observed and clinically used in the field of radiotherapy.**

# Introduction

Bystander effect as a cellular response to ionizing radiation was firstly described many years ago (Nagasawa and Little, 1999) and it can be defined as the process in which cells that have not been directly exposed to ionizing radiation, show the same DNA instability, and eventually cell death, than those exposed cells (Fig.1) (Little, 2003) (Sowa, 2005). Since the first reports of this effect from the field of radiology, the nature of the interactions that could produce these responses has been broadly investigated(Mykyta V. Sokolov, 2010, Rajalakshmi S. Asur, 2009, C Shao, 2008, Y Jiang, 2014). Although many open issues remain unclear about the kind of cell communication, there are solid basis pointing to chemical signalling as the main mechanism involved in the transmission of information from the exposed cells to neighbouring ones(Hamada et al., 2007). There is evidence that oxygen reactive species and secreted factors play a key role in the induction of the bystander effect(Azzam et al., 2003). It is also well known that the damage induced in the cells not exposed to ionizing radiation by the bystander effect it is not necessary the same as the induced in cells directly irradiated(Ward, 2002). These indirect mechanisms constitute the basis of the bystander effect and are relevant at low dose radiation exposure being saturated at high doses(Prise and O'Sullivan, 2009). Some studies have also demonstrated that tumour cells are more sensitive than healthy cells to the bystander effect coming from irradiated cells, resulting in an advantage in tumour treatment(Jaime Gómez-Millán, 2012). Mothershill and Seymour put forward clear evidence of the existence of this indirect effect observing that when the medium in which the cells were irradiated is added to an non-irradiated cell culture, the same level of cell death occurs respect to the culture submitted to irradiation(Mothersill C, 1997). Several mediating mechanism have been proposed for bystander effect (Hamada et al., 2007). Direct irradiated cells can transmit the bystander effect to neighbouring cells by gap junctional intercellular communication or by releasing soluble species (such as ROS, NO, $Ca^{+2}$ etc.) into the medium (Seymour, 1988, Little, 2006, Shu Guo 2014, Baskar, 2010, E-C Liao, 2014). It has been recently proposed that bystander signalling between cells can be also mediated by exosomes(Andrea Sobo-Vujanovic a, 2014)' (K. Kumar Jella, 2014).

Hyperthermia is a cancer therapy protocol based on the overheating of cancerous target tissues above physiological temperatures (41-46 ºC) to eliminate the malignant cells at that region. Hyperthermia can be used as a standalone therapy or as a synergistic therapy with radiotherapy, allowing a reduction of radiated doses(Sannazzari et al., 1989). A recent nanotechnology-based approach has been introduced in the clinic as a focused hyperthermia based in the tumour ablation using magnetic nanoparticles (MNPs). This therapy propose the use of single-domain magnetic nanoparticles MNPs as heating agents, through the application of a low-radiofrequency magnetic field (100 kHz $< f <$ 800 kHz) and named magnetic hyperthermia (MHT) (Jordan et al., 2009). The basic mechanism for heat production is related to the coupling of the magnetic moment of the MNPs and the external alternative magnetic field (AMF). The efficiency of the MNPs to absorb energy from the (AMF) depends on the size, size distribution and magnetic anisotropy of the MNPs and the amplitude and frequency of the AMF (Asin L., 2012) .

The electromagnetic radiation used for MHT is in the frequency range of the AM broadcasting radio frequency, at these frequencies and AMF amplitudes used in the MHT experiments, the radiation dose poses no harm to the human body since they are almost non-interacting with organic matter. An overwhelming activity in the field of nanomedicine has established in vitro experiments, in which the cell cultured with MNPs gives rise to their uptake and internalization in the cellular media. Under further exposure of the cell culture to AMF a temperature rise take place and the cells cannot survive above a determined value (typically 46ºC), this is the general frame of cellular MHT. Nevertheless, recent experiments have shown that tuning MHT conditions, massive cell death can occur without any detectable macroscopic temperature increase in the cell culture medium (Cheng et al.). The mechanisms for the cell dead without increasing the medium temperature are not clear so far.

Up to now, the bystander effect term has been exclusively used in the field of ionizing radiation; however previous studies have demonstrated that supernatants in MHT experiments are toxic for cells that were never exposed to radiofrequency radiation (Asin et al., 2013). As a consequence we propose that both ionizing and non-ionizing radiations trigger damage in not directly exposed cells and consequently the term bystander effect can be also used in the field of MHT (Fig. 1). As in the case of ionizing radiation, the causes that provoke damage in surrounding cells have not been identified yet. Here we show that there are some factors, as magnetic nanoparticles (MNPs) location and radiation-MNPs interaction that causes cell death. Under determined circumstances the collected medium can induce cell death following bystander effect mechanism [24], as it was found in the case of ionizing radiation (Mothersill C, 1997). In order to understand the origin of the bystander effect in cellular MHT, we have investigated the influence of final biodistribution of MNPs in two different kinds of cell lines: dendritic cells (DCs) and J774 macrophage cells following the procedure described in Method section. The experimental evidence of the toxicity of the supernatant (medium) after MHT-treated cells culture, when added to non-treated cells has been further investigated in DCs and J774 cells by using different types of MNPs. The results indicate that the observed bystander effect occurs only when the temperature does not rise after AMF application and the MNPs are located within the cell lysosomes. Nevertheless, it is absent when cell death is triggered by temperature rising of the cell medium without irradiation. As consequence we can infer that under AMF exposure, MNPs can induce a leaking of the lysosome content that in turn provoke a radically different biological consequences respect to the case of randomized MNPs distribution at endosomes or at the cell membrane.

Here we report the results of the "in-vitro" investigation of cell death by MHT following the temperature of the cell culture and considering the MNP allocation in the cells. In addition we also investigated the cell death by increasing the temperature of the cell culture with MNP, in absence of AMF irradiation.

## Materials and Methods

**Cell culture**

For DCs culture, peripheral Blood Mononuclear Cells (PBMCs) were obtained from healthy blood donors by Ficoll density gradient (Ficoll Histopaque-1077 Sigma). Cells were washed twice with PBS for 7 minutes at 1200 rpm at room temperature. To take away platelets, cells were centrifuged 10 minutes at 800 rpm. Monocytes (CD14+) were isolated with immunomagnetic beads (CD14 Microbeads, Miltenyi) by positive immunoselection using the autoMACS Separator (Miltenyi) as described by the manufacturer.

In T75 flask, purified monocytes (10⁶/ml) were cultured in RPMI 1640 (Sigma) with 10% FBS, 1% glutamine, 1% antibiotics (100 U/ml penicillin and 100 ng/ml streptomycin) and supplemented with IL-4 (25 ng/ml) and GM-CSF (25 ng/ml)(Sigma). Cells were performed at 37ºC in humidified atmosphere containing 5% $CO_2$ for 5 days. Every 2 days, medium was replaced by fresh medium containing the same concentration of interleukins.

J774 cells were cultured in DEMEM supplemented with 10% FBS, 1% glutamine, 1% antibiotics (100 U/ml penicillin and 100 ng/ml streptomycin). Twice a week cells were passed by tripsinization.

**Magnetic and fluorescence nanoparticles**

Fluorescent particles (micromer®-redF) were composed of a polysteryrene core with carboxyl (COOH-) groups at the surface, having hydrodynamic radio of 250nm as measured by dynamic light scattering. These NPs have rhodamine fluorophore at the surface, and are dispersed in a water-based carrier liquid at a concentration of 25 mg particles/mL (labelled Rho-NPs). For the PEI-coated $Fe_3O_4$ MNPs (labelled PEI-MNPs), the synthesis was based on a modified hydrolysis route based on precipitation of FeSO4 in NaOH with a mild oxidant, already reported elsewhere.[8] This route allowed to control the particle size by in situ adding a functionalizing polymer (polyethyleneimine PEI, 25 kDa). The free amine groups of PEI-MNPs were tagged with the fluorescent dye alexa 488(TFP)(Invitrogen). 2.5 mg of PEI-MNPs were suspended in 1 ml of 0.1 M sodium carbonate-bicarbonate buffer at pH 9.5 to ensure the deprotonation of amine groups on the PEI polymer coating. Alexa 488 solution (1 mg/ml DMSO: 20 μl) was then added to the PEI-MNPs suspension and the reaction mixture suspension was covered with aluminum foil and rotated (using a Rotator) at room temperature for 3 hours. The Alexa 488-labelled PEI-MNPs were finally washed with deionized $H_2O$ until no further fluorescence was observed in the supernatant. The sample is denoted as f-PEI-MNPs. Commercial MNPs (nanomag®-D from Micromod GmbH) were used for magnetic hyperthermia experiments carried out on DCs. These MNPs were composed of a magnetic core ($Fe_3O_4$) functionalized with dextran

(carboxylic groups at the surface) having a hydrodynamic diameter of 250nm (labelled COOH MNPs).

**Confocal microscopy**

Co-localization study to analyse the final fate of the NPs within the cells was performed by an in vivo staining using Lysotracker® Green (DCs) and Red (Invitrogen) (J774 cells). DCs, at day five, and J774 cells were placed into each 60 µ-dish ibiTreat (Ibidi GmbH) and 50µg/ml of fluorescent Rho-NPs were added. J774 cells were incubated with 50µg/ml of f-PEI-MNPs .After an overnight incubation cells were washed three times with fresh medium and were incubated with 75nM of Lysotracker for 15 minutes. Cells were then washed again twice with fresh medium and observed under an Olympus compact confocal microscope. All the samples were analysed under the same settings conditions.

**Transmission electron microscopy**

On day 5 of culture, $3 \times 10^6$ cells/well were seed into 12-well-plate in 2mL of medium supplemented with cytokines. DCs were incubated with 50ug/ml of COOH MNPs. The samples were incubated overnight at 37ºC. The following day 2mL of Glutaraldheyde 4% in sodium cacodylate 0,2M pH=7,2 were added to each well and the plate was incubated 2h at 4ºC. Next, cells were collected, spun down, resuspended in 1mL of Glutaraldheyde 2% in sodium cacodylate 0,1M and keep at 4ºC.

Then cells were washed three times with sodium cacodylate 0,1M and fixed with 250 µl of potassium ferrocianide 2,5 % in sodium cacodylate 0,1M and 250 µl of osmium tetraoxide 2 % 1h at room temperature keeping from the light. After that, cells were washed twice with cacodylate 0,1M and the dehydratation process was performed resuspending the cells in increasing concentrations of acetone (50, 70, 90 and 100% 10 minutes each twice).

Cells were infiltrated overnight in a shaker at room temperature with a 1:1 EPON-acetone 100% mixture and the following day this mixture was replaced by 100% EPON resin and incubated for 5 hours in a shaker at room temperature. After that, cells were pelleted in pure EPPON resin and baked at 60ºC for 48h. Ultrathin sections (60-80nm) placed onto a cupper grid treated with 2% Uranyl acetate in water for 45 minutes at room temperature, after that, grids were gently washed with distillate water and treated with 2% lead citrate in water for 5 minutes in presence of NaOH in order to maintain a desiccate atmosphere. Samples were observed in the Unitát Microscopia Electrónica (Universidad Barcelona) a transmission electron microscopy Jeol EM 101 microscope at an accelerating voltage of 80 KV.

**Alternating magnetic field experiments**

The exposure of the cells to an AMF was performed with a commercial ac applicator (model DM100 by nB nanoscale Biomagnetics, Spain) working at f = 580 kHz and field amplitude of 300Oe. The applicator is equipped with an adiabatic sample space (~ 0.5 ml) for measurements in liquid phase. Cells were cultured overnight at 37ºC with 100ug/mL of PEI-MNPs and 100 μg/mL of COOH MNPs. The following day cells were washed, collected and resuspended in 500 μl of complete medium. Each sample consisted of $10^7$ cells.

# Results

First, we will focus on the biodistribution of the MNPs within the cells. The internalization mechanisms of MNPs follow the usual pathway of endocytic vesicles. To elucidate if the location of the MNPs is in the lysosome, we performed an *in vitro* cellular labeling with Lysotracker, which is a pH probe that exhibit either green or red fluorescence only at low pH (typical for lysosomes) indicating the lysosome location in the cytoplasm. To track the nanoparticles either in J774 or in DCs cells, we used fluorescent green (Alexa TFP) dyes (f-PEI-MNPs and red (micromer®-redF) Rho-NPs) respectively. Confocal microscopy allows obtaining independent images filtered in wave length that show the different region occupied by the lysosomes and the MNPs respectively. The overlaying of the images allows a precise determination of the co-localization or not of the MNPs in the lysosome space (see Fig. 2).

The images of DCs show the co-localization of the MNPs in the lysosomes, as the overlay of the two channels shows that the internalized MNPs have the same bio-distribution as the lysosomes (see Fig. 2, upper row and Methods section). Similar procedure was followed with a macrophage cell line J774 and confocal microscopy showed (see Fig. 2, lower row) that distribution of the MNPs does not correspond to lysosomes. MNPs are located in the cytoplasmatic region but they do not co-localize with lysosomes. The localization of MNPs within DCs cells was confirmed by transmission electron microscopy (TEM) (see Fig.3). It was found that the nanoparticles were highly concentrated at endocytic vesicles previously characterized previously as lysosomes by confocal microscopy.

Second, we performed experiment of MHT in both cell lines in the case of co-localization (dendritic) and no co-localization (macrophages J774) (see Fig. 4). MNPs loaded DCs were exposed to radiofrequency at tuned parameter (Methods section), which induced the cell death without increase of the medium temperature. The resultant supernatant was collected and added to a cell culture without MNPs and no irradiated (as sketched in Fig. 1). The results clearly show the cell dead is induced by the toxic bystander factors. However, the application of radiofrequency in the case of the J774 cell, in which no co-localization was observed, cell dead need to be induced by rising the temperature of the medium up to 60 ºC under AMF application (MHT) and no bystander effect was observed. This was confirmed by the cell viability after exposure of pristine cell culture to the supernatant obtained in the MHT experiment. We can argue that due to the waste cellular material contained at the lysosomes, if the MNPs inside these organelles can deliver energy during the AMF application, it may cause the disruption of the lysosome membrane. The release of the lysosome content to the cytoplasm acts as bystander factor, being toxic to the DCs and causing cell death of unloaded cell. The same argument applies for the pristine DC culture when exposed to the toxic supernatant obtained from MHT experiment(Asin L., 2012). The ability of MNPs to destabilize the lysosome membrane during the application of AMF has been demonstrated by Domenech *et al*. (Maribella Domenech, 2013). A possible mechanism could be local heat delivery at the lysosome space. Recently, it has been proposed that, in an inhomogeneous magnetic field of low frequency, MNPs could generate ultrasound waves(Carrey et al., 2013). We argue that this last mechanism could be also responsible for the disruption of the lysosome membrane. If this is the case, MHT could be of major relevance as source of focalized ultrasound therapy.

Third we study in both cell lines the effect of increase of the medium temperature by normal heating without AMF exposure, i.e., in the absence of applied radiation. The results indicated that cell death pathway undergone in water bath heated cells does not depend on the MNPs content. Cell death was observed at temperatures of 50º and 60ºC and the obtained supernatants were no toxics for untreated cell cultures(data not shown). These results were found to be independent if the MNPs colocalize or not in lysosomes.

## Discussion

We can conclude that bystander effect occurs in MHT when radiofrequency fields are applied to MNPs loaded cells in which the nanoparticles colocalize in lysosomes. In this case bystander factors induce indirect cellular dead. This effect is observed at the threshold of several parameters, requiring critical tuning of concentration of loaded nanoparticles, location in the cellular medium, intensity of the applied radiofrequency field and time exposure; in such a way that the MNPs loaded cells died without increasing the medium temperature.

We propose that the parallel observation of bystander effect by ionizing radiation on cells and radiofrequency AMF applied on MNPs loaded cells, points to a common origin which could interconnect scientific problems in both scientific communities and may give rise a rapid advance in the understanding of the mechanisms of the bystander effect at cellular level. As this effect is considered to play a relevant role in the tumour regression in radiotherapy our contribution might be of fundamental interest in order to progressively incorporate nanotechnology through the implant of MNPs as a key beneficial ionizing radiation free therapy.

## Acknowledgements

We are grateful to the IACS (University Hospital, Zaragoza) and to the Advance Microscopy Laboratory-INA staff for their advice and technical support. Dr. C. Tortoglione and Dr. J. Carrey for clarifying discussions.

## Declaration of Interest Statement

This work was supported by the Spanish Ministerio de Economía y Competitividad (project MINECO MAT2010-19326). LA acknowledges MINECO by financial support through a FPU fellowship. The authors report no conflicts of interest.

# Figures

**Figure 1.**

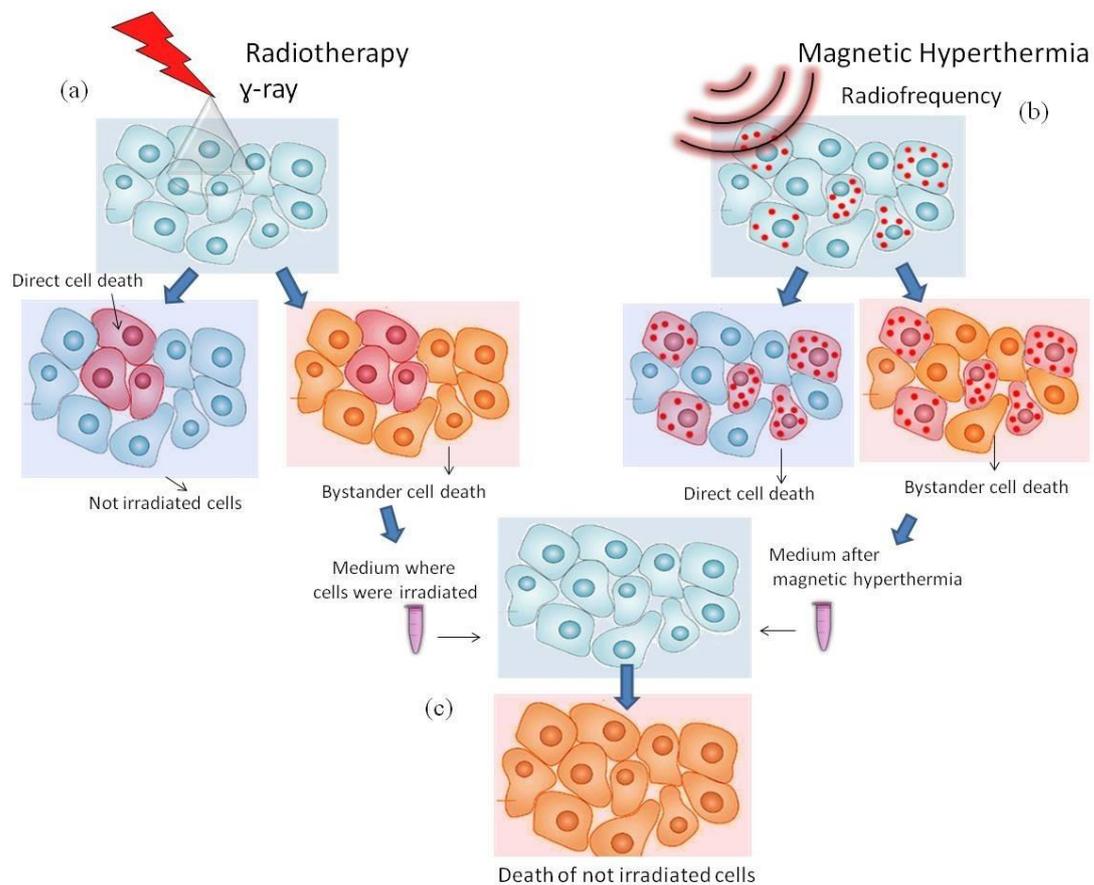

**Figure 2.**

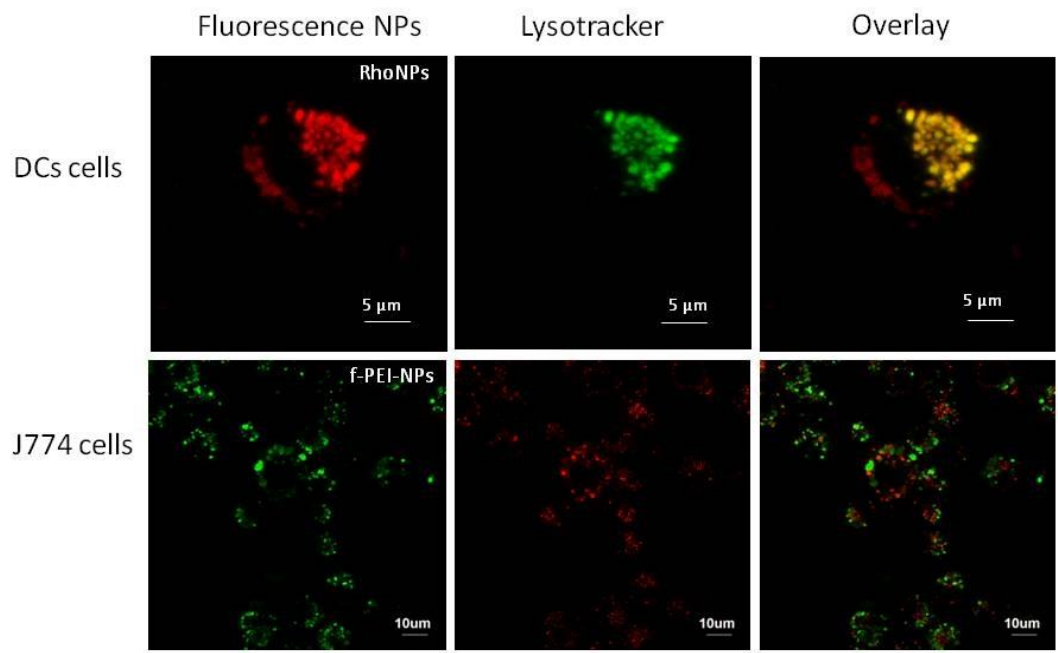

**Figure 3.**

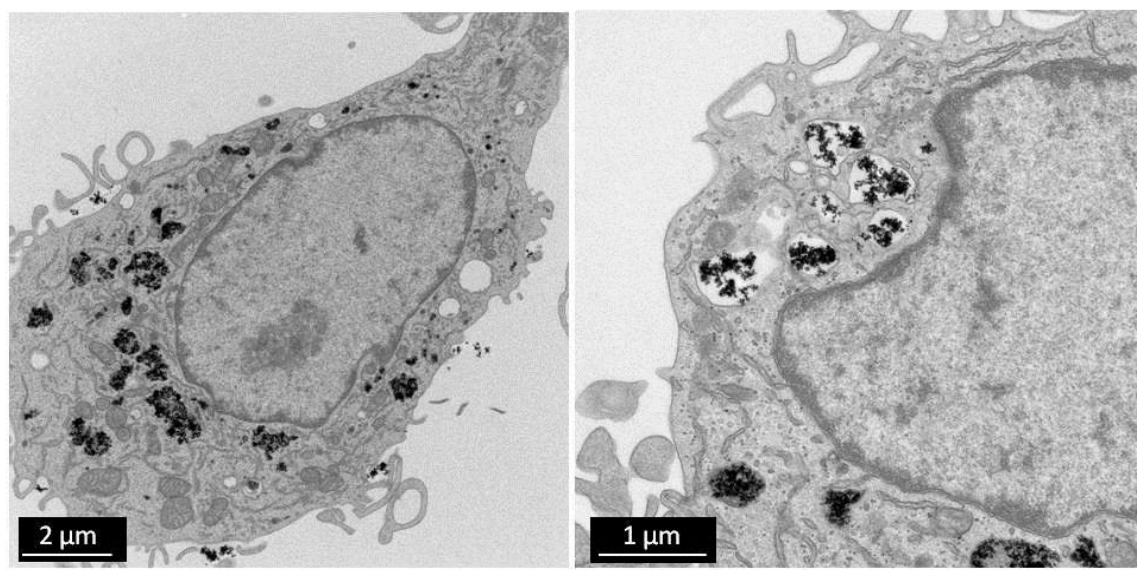

**Figure 4.**

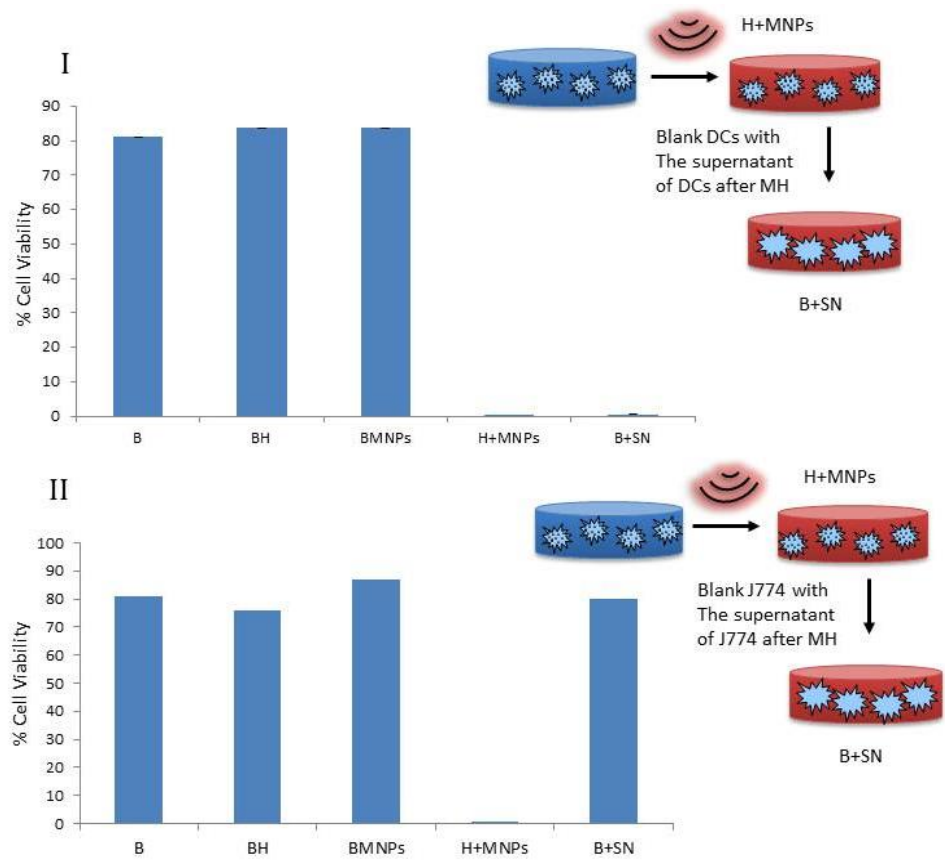